\documentclass[]{spie}  
\usepackage[]{graphicx}
\usepackage{epstopdf}
\usepackage{amsfonts}
\usepackage{amsmath}
\usepackage{amssymb}

\usepackage{hyperref}
\newcommand{\bra}[1]{\ensuremath{\left\langle#1\right|}}
\newcommand{\ket}[1]{\ensuremath{\left|#1\right\rangle}}

\title{Superconducting circuitry for quantum electromechanical systems} 

\author{Matthew D. LaHaye\supit{a}, Francisco Rouxinol\supit{a}, Yu Hao\supit{a}, Seung-Bo Shim\supit{b}, Elinor K. Irish\supit{c}
\skiplinehalf
\supit{a}Department of Physics, Syracuse University, Syracuse, New York 13244-1130, USA; \\
\supit{b}Korea Research Institute of Standards and Science, Daejeon, 305-340, Republic of Korea;\\
\supit{c}Physics and Astronomy, University of Southampton, Highfield, Southampton, SO17 1BJ, United Kingdom
}

\authorinfo{Further author information: (Send correspondence to M.D.L.)\\M.D.L.: E-mail: mlahaye@syr.edu, Telephone: 1 315 443 2564}

\begin{document} 
  \maketitle 

\begin{abstract}
Superconducting systems have a long history of use in experiments that push the frontiers of mechanical sensing.  This includes both applied and fundamental research, which at present day ranges from quantum computing research and efforts to explore Planck-scale physics to fundamental studies on the nature of motion and the quantum limits on our ability to measure it.  In this paper, we first provide a short history of the role of superconducting circuitry and devices in mechanical sensing, focusing primarily on efforts in the last decade to push the study of quantum mechanics to include motion on the scale of human-made structures.  This background sets the stage for the remainder of the paper, which focuses on the development of quantum electromechanical systems (QEMS) that incorporate superconducting quantum bits (qubits), superconducting transmission line resonators and flexural nanomechanical elements. In addition to providing the motivation and relevant background on the physical behavior of these systems, we discuss our recent efforts to develop a particular type of QEMS that is based upon the Cooper-pair box (CPB) and superconducting coplanar waveguide (CPW) cavities, a system which has the potential to serve as a testbed for studying the quantum properties of motion in engineered systems.   
\end{abstract}


\keywords{quantum electromechanical systems, nanomechanics, superconducting qubits, hybrid quantum systems}


\section{Introduction} 

This section first provides some historical background on the use of superconducting circuitry for mechanical detection in order to trace the origins and establish the present day context of the main focus of this paper: namely, QEMS that incorporate superconducting qubits, cavities and nanomechanical devices.  It then outlines the basic model for a particular type of QEMS: CPBs coupled to nanomechanical resonators.  In the end, the state-of-the-art for this system is reviewed and current experimental challenges are discussed. It should be noted that the introductory review does not attempt to do justice to the parallel (and increasingly interdependent) field of optomechanics.  For more information on that field, several excellent reviews are cited below, which we recommend to interested readers.  

\subsection{The Origins of Quantum Electromechanical Systems} 
\label{sec:title}

The use of superconducting systems for sensitive measurements of motion traces back at least 50 years to the origins of resonant-mass gravitational-wave (GW) antennas\cite{weber1960detection,braginsky1985systems,blair1995high,harry2000two,marin2013gravitational} .   Over the decades, superconducting technology has played an integral role in that field, with massive, cryogenically-cooled superconducting bars serving as the high-Q acoustic cavities at the heart of the GW antennas, and superconducting quantum interference devices (SQUIDs)\cite{harry2000two,marin2013gravitational} or superconducting microwave resonators\cite{blair1995high} serving as ultrasensitive front-end detectors in the transducer circuitry.    

A parallel track in the history of superconducting devices and mechanical detection arose in the 1990’s, with the emergence of nanomechanics\cite{cleland1996fabrication} and the recognition that nanoelectromechanical systems (NEMS)\cite{roukes2000nanoelectromechanical,ekinci2005nanoelectromechanical} could serve as a new frontier for studying macroscopic quantum effects.\cite{cleland1999nanoscale,roukes2000nanoelectromechanical,blencowe2000quantum,roukes2001nanoelectromechanical,blencowe2000sensitivity,schwab2001quantum,milburn2001quantum,zhang2002intrinsic,armour2002mechanical,armour2002entanglement,irish2003quantum} Indeed SQUID-based detection and superconducting bias circuitry integrated with nanomechanical structures enabled the first measurements of the quantum of thermal conductance in 1999\cite{schwab2000measurement}.  Moreover, researchers at the time, inspired in part by earlier developments in the GW-detection community\cite{braginsky1995quantum,caves1980measurement,bocko1996measurement}, also realized that full exploration of quantum NEMS (or more generally QEMS) would require the development of new detectors and control circuitry that could be integrated strongly with motional degrees of freedom at the nanoscale and yet simultaneously provide unprecedented resolution and minimal back-action, themselves operating in regimes governed by quantum mechanics\cite{roukes2000nanoelectromechanical,blencowe2000quantum,roukes2001nanoelectromechanical,blencowe2000sensitivity,schwab2001quantum,milburn2001quantum,zhang2002intrinsic,armour2002mechanical,armour2002entanglement,irish2003quantum,cleland2004superconducting,blencowe2004quantum,schwab2005putting}.  Crucially, natural solutions to these challenges emerged from the nascent field of superconducting quantum computation. 

During the 1990’s a variety of mesoscopic superconducting devices were developed\cite{schoelkopf1998radio,nakamura1999coherent,devoret2000amplifying,Makhlin01} that became important candidates in the next decade as both detector elements and quantum bits (qubits) in scalable quantum processing architectures.\cite{Makhlin01,schoelkopf2008wiring,clarke2008superconducting,devoret2013superconducting} In these systems, at milli-Kelvin temperatures, the interplay of charging and Josephson effects\cite{fulton1989observation,van1991combined,michael2004introduction} can give rise to noise characteristics dominated by quantum transport processes\cite{clerk2002resonant,clerk2005quantum,blencowe2005dynamics,xue2009measurement,clerk2010introduction} and, in properly tuned devices, quantum coherent behavior\cite{nakamura1999coherent,Makhlin01,koch2007charge,schreier2008suppressing,schoelkopf2008wiring,clarke2008superconducting,devoret2013superconducting} analogous to that seen in atomic and spin-based systems.   The same properties also make these devices ideally suited for sensing and controlling the quantum properties of mechanics.\cite{blencowe2000sensitivity,schwab2001quantum,milburn2001quantum,zhang2002intrinsic,armour2002mechanical,armour2002entanglement,irish2003quantum,cleland2004superconducting,blencowe2004quantum,clerk2005quantum,blencowe2005dynamics} Moreover, their size scale and material composition are commensurate with typical nano- and micromechanical systems, enabling the use of standard fabrication processes to engineer the devices \textit{on chip} with the mechanical elements, in order to achieve precisely controlled and even tunable interactions between the systems.\cite{knobel2003nanometre,lahaye2004approaching,naik2006cooling,flowers2007intrinsic,etaki2008motion,lahaye2009nanomechanical,o2010quantum,suh2010parametric,pirkkalainen2013hybrid}   

An early example of this synergy was seen with the single-electron transistor (SET) and its superconducting cousin the SSET.\cite{schoelkopf1998radio,devoret2000amplifying,fulton1989observation,van1991combined,michael2004introduction} By the late 1990's the SET was recognized as a potentially quantum-limited electrometer, with sufficient bandwidth, when operated in microwave circuitry (RF-SET), to perform single-shot quantum-state detection of charge-based qubits.\cite{schoelkopf1998radio,devoret2000amplifying,lehnert2003measurement} Soon thereafter it was appreciated that the unprecedented charge sensitivity ($\sim \mu \rm{e}/\sqrt{\rm{Hz}}$) and large bandwidth ($\sim100\,{\rm MHz}$) could also be utilized for performing continuous, linear displacement detection of MHz-range nanomechanical elements, with sensitivity approaching the limit allowed by the Heisenberg Uncertainty Principle.\cite{blencowe2000sensitivity,zhang2002intrinsic} This motivated several experimental efforts to integrate MHz-range NEMS with linear displacement transducers based upon SETs\cite{knobel2003nanometre} and RF-SSETs.\cite{lahaye2004approaching} Subsequent theoretical\cite{clerk2005quantum,blencowe2005dynamics} and experimental development\cite{naik2006cooling} of the SSET displacement detector in fact showed the coupled SSET-NEMS device to be a system with rich dynamics: it allowed for displacement detection near the uncertainty principle limit at particular SSET Cooper-pair/quasiparticle transport resonances; provided the first demonstrations of the quantum back-action of fundamental particles on the motion of a macroscopic mechanical system; and enabled detection of nanomechanical motion for the first time at low thermal occupation numbers, where observation of quantum effects in the behavior of the mechanics might reasonably be expected.

In the early 2000’s it was also appreciated that coherent superconducting devices like the Cooper-pair box (CPB)\cite{nakamura1999coherent,Makhlin01} and the phase qubit\cite{martinis2002rabi} could be utilized to go beyond linear displacement detection and to enable the capability to manipulate and measure patently quantum mechanical states of nano- and micromechanical modes,\cite{armour2002mechanical,armour2002entanglement,irish2003quantum,cleland2004superconducting}  in analogy to systems in cavity quantum electrodynamics (CQED) and ion-trap physics that had enabled groundbreaking research on the quantum properties of light and trapped ions.\cite{haroche2006exploring} Initial theoretical proposals put forth in the literature posited Josephson-junction-based qubits as tools for performing a variety of tasks, including the measurement and preparation of nanomechanical superposition states, number states and zero-point energy;\cite{armour2002mechanical,armour2002entanglement,irish2003quantum} as well, protocols were outlined for use of qubit-coupled nanoresonators (QCNR) as quantum memory and bus elements.\cite{cleland2004superconducting} These initial proposals and the explosion in subsequent years of new superconducting qubit technology, most notably circuit QED (cQED) architectures based upon superconducting transmission line resonators,\cite{koch2007charge,schreier2008suppressing,schoelkopf2008wiring,clarke2008superconducting,devoret2013superconducting,wallraff2004strong} fueled a myriad of proposals\cite{martin2004ground,rabl2004generation,wei2006probing,tian2006scheme,clerk2007using,jacobs2007continuous,utami2008entanglement,armour2008probing,jacobs2008energy,semiao2009kerr,heikkila2014enhancing} over the ensuing decade for the sake of exploring fundamental aspects of quantum mechanics such as the quantum-to-classical transition, the fundamental limits to the sensing of motion, and applications in quantum information processing and metrology. As well, it has since been appreciated that these systems offer the potential to study new regimes of the paradigmatic Jaynes-Cummings model\cite{haroche2006exploring}, going beyond the rotating-wave approximation.\cite{irish2005dynamics,zueco2009qubit}

Notwithstanding the extensive theoretical effort to develop QCNRs, progress on the experimental front has been slow due to an array of challenges that will be elaborated in Section 1.3.  Nonetheless, there have been several important developments in the field beginning in 2009 with the first demonstration of the interactions between a nanomechanical flexural resonator and a superconducting charge qubit.\cite{lahaye2009nanomechanical}  The experiment in 2009 demonstrated that, for a CPB and nanoresonator whose energies were far out of resonance, a simple electrostatic interaction between the systems gives rise to shifts in the energy of the nanoresonator that are dependent on the qubit's state.  Such dispersive shifts are analogous to single-atom index effects observed in some CQED systems\cite{haroche2006exploring} and in principle could be utilized for a multitude of tasks if developed further, including for generating highly-non-classical  states of mechanics \cite{utami2008entanglement,armour2008probing,suh2010parametric,semiao2009kerr,jacobs2009engineering} and for generating a quantum switch to shuttle information coherently between multiple mechanical modes\cite{mariantoni2008two}.  In 2010, shortly after these initial results, results were put forth demonstrating the first use of a superconducting qubit to manipulate and measure the quantum properties of a mechanical device.\cite{o2010quantum}   In this work, a micromechanical piezo-disk resonator was integrated with a superconducting phase qubit; sophisticated techniques that had been developed for controlling and measuring the phase qubit were then adopted to perform quantum Rabi swapping and Ramsey interference experiments with the micromechanical mode.  This experiment was a milestone not only for the field of mechanical quantum systems, but for the entire physics community, providing the first demonstration of energy quantization and quantum superposition states with a normally-classical macroscopic mechanical mode.  More recently, in 2013, observations of dispersive interactions between a transmon qubit and micromechanical drumhead that were complimentary to the results in 2009 were published.\cite{pirkkalainen2013hybrid} Specifically, these results showed evidence for mechanical Stark shifts in the transmon's energy spectrum; the shifts were shown to be proportional to the number of quanta in the mechanical mode and thus analogous to the traditional AC Stark shift\cite{haroche2006exploring} seen in atomic physics, CQED and cQED. While single-quantum shifts were not resolved in the 2013 work, the capability is within reach using current technology (as discussed in Section 2) and could enable projective measurements and even quantum non-demolition (QND) measurements of the energy of nano- and microscale resonators.\cite{irish2003quantum,clerk2007using} Such techniques would find myriad applications in areas ranging from quantum information processing to the study of quantum fluctuation theorems\cite{campisi2011colloquium,brito2014testing} and fundamental investigations of how energy is transported and dissipated in nanoscale devices.        

Alongside the development of QCNRs, intense effort has been directed toward an additional branch of superconducting electromechanical systems: microwave cavity mechanics\cite{regal2008measuring,hertzberg2010back,rocheleau2010preparation,massel2011microwave,teufel2011sideband,palomaki2013coherent,palomaki2013entangling,suh2014mechanically}.  While a full accounting of the origin of these systems is beyond the scope of this historical introduction, it is fair to say that they catalyzed from (and amidst) a confluence of diverse research directions including prior pioneering work on dynamical back-action in the GW community\cite{blair1995high,braginsky1995quantum} and contemporaneous research in the mid-2000’s in the fields of cQED\cite{wallraff2004strong}, superconducting astrophysical detectors\cite{day2003broadband}, and optomechanics\cite{kippenberg2008cavity,aspelmeyer2012quantum,aspelmeyer2014cavity}. By and large the cavities at the heart of these systems have been high-quality superconducting circuit resonators that have been engineered to provide parametric read-out and control of flexural type nano- and micromechanical modes.  The earliest versions of microwave cavity mechanics utilized transmission-line resonators, primarily in coplanar waveguide (CPW) geometries.\cite{regal2008measuring,hertzberg2010back,rocheleau2010preparation} However the most successful, and now most widely used, scheme involves the integration of a micromechanical membrane structure as one electrode of a parallel-pate capacitor in a lumped-element LC circuit.\cite{teufel2011sideband,palomaki2013coherent,palomaki2013entangling,suh2014mechanically}  The parametric coupling that can be achieved between membrane modes and the LC circuit in this configuration is several orders of magnitude greater than what has been demonstrated using CPW geometries and has enabled the use of side-band-resolved driving\cite{kippenberg2008cavity,aspelmeyer2014cavity,poot2012mechanical} of microwave cavities for a growing list of accomplishments:  cooling of a MHz-range micromechanical mode to its quantum grounds state,\cite{teufel2011sideband} a feat not yet achieved using passive cryogenic refrigeration; coherently storing and retrieving quantum states of microwave fields in a mechanical mode;\cite{palomaki2013coherent} generating and characterizing entanglement between the motion of a mechanical mode and the electric field of a traveling microwave signal;\cite{palomaki2013entangling} and detecting, as well as partially evading, the quantum back-action noise of a microwave field in the measurements of mechanical motion.\cite{suh2014mechanically} The potential applications of these spectacular advances are numerous and range from the use of cavity-cooled mechanics to generate complex entangled states for teleportation and entanglement-swapping protocols, quantum squeezed states of motion for the detection of weak forces, and fundamental explorations of quantum mechanics in new limits.\cite{aspelmeyer2012quantum,poot2012mechanical,aspelmeyer2014cavity}  

The evolution highlighted here continues at the time of writing, with the parallel tracks noted above intermixing as well as incorporating new devices and materials.  Hybrid quantum systems\cite{xiang2013hybrid} in a multitude of forms are being developed that either currently incorporate or ultimately will require superconducting QEMS:   microwave-to-optical mechanical transducers in order to coherently link these disparate energy scales in future quantum networks;\cite{hill2012coherent,bochmann2013nanomechanical,bagci2014optical,andrews2014bidirectional} superfluid cavity mechanics, ultra-high-Q systems that incorporate superfluid acoustic modes within a 3D microwave cavity;\cite{de2014superfluid}  cavity mechanics that integrate novel mechanical elements such as carbon nanotube resonators and suspended graphene sheets with superconducting cavities;\cite{singh2014optomechanical,weber2014coupling} and surface acoustic wave (SAW) circuits resonantly interfaced with transmon-type qubits,\cite{gustafsson2014propagating} to name a few.  

For many of these hybrid QEMS, including all of the ones mentioned above, further development of QCNR would have direct relevance for their future applications - if for no other reason than to utilize the QCNR as a tool for generating and detecting highly-nonclassical states of the mechanical components. Thus the remainder of the paper will focus on discussing some of the challenges facing further experimental development of the QCNR, particularly the CPB-based version, and the efforts by the authors to overcome these challenges.

\subsection{Canonical Model for the Cooper-Pair Box and Nanomechanical Resonator} 
\begin{figure}
\begin{center}\includegraphics[ width=1\columnwidth,keepaspectratio]{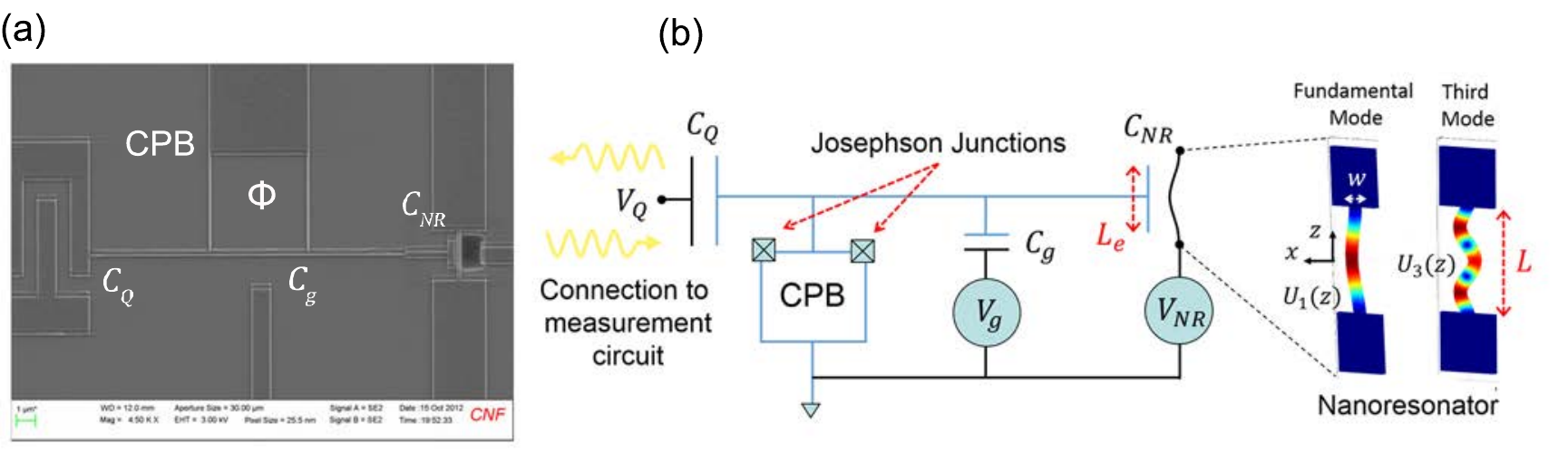}\end{center}
	\caption{(a) SEM micrograph of the first generation of CPB-type QCNR fabricated and measured by the authors. The CPB and nanostructure are patterned out of aluminum atop a high-resitivity silicon substrate. Additional details of the measurement process and the device are discussed in Section 2. (b) Basic circuit schematic for the device in (a) and mode shapes $U_1(z)$ and $U_3(z)$ for the fundamental mode and third mode respectively. Note the location of the CPB is flipped from its position in (a) in order to simplify the schematic. The second mode of the nanostructure is not illustrated; due to the asymmetry of the mode with respect to the CPB electrode, its motion should couple negligibly to CPB charge. Also note that the thickness parameter $t$ is not defined in the illustration, but is simply the out-of-plane thickness (in the $y$ direction) of the structure. }
	\label{fig:fig1}
\end{figure}
Devices like the CPB-based QCNR shown in Figure 1 are typically modeled in the literature using different limits of the following Hamiltonian\cite{irish2003quantum,lahaye2009nanomechanical}
\begin{equation}
\hat{H}=\hat{H}_{\mathit{CPB}}+\hat{H}_{\mathit{\mathit{\mathit{NR}}}}+\hat{H}_{\mathit{INT}},
\label{hamiltonian}
\end{equation}
which is composed of a contribution from the CPB that is given by 
\begin{equation}
\label{HamiltonianCPB}
\hat{H}_{\mathit{CPB}}=4E_C\sum_n(n-n_\Sigma)^2\ket{n}\bra{n}-\sum_n\left[\frac{{\cal E}_J(\Phi)}{2}\ket{n}\bra{n+1}+\frac{{\cal E}_J^\ast(\Phi)}{2}\ket{n+1}\bra{n}\right],
\end{equation}
a component due to the nanoresonator
\begin{equation}
\hat{H}_{\mathit{\mathit{NR}}}=\hbar\omega_{\mathit{NR}}(\hat{a}^{\dagger}\hat{a}+\frac{1}{2}),
\label{HamiltonianNR}
\end{equation}
and a term representing the electrostatic interaction between the systems 
\begin{equation}
\hat{H}_{\mathit{INT}}=\hbar\lambda\sum_n(n-n_{\Sigma})\ket{n}\bra{n}(\hat{a}^{\dagger}+\hat{a}).
\label{HamiltonianINT}
\end{equation}
In Eq.(\ref{HamiltonianCPB}), $E_C$ and $E_J(\Phi)$ are the charging and Josephson energies of the CPB respectively. The value of $E_C$ sets the scale for the CPB's electrostatic energy, which can be tuned by adjusting the polarization charge $n_\Sigma=C_QV_Q/2e+C_gV_g/2e+C_{\mathit{NR}}V_{\mathit{NR}}/2e$ on nearby electrodes, where the capacitances $C_Q$, $C_g$ and $C_{\mathit{NR}}$ and voltages $V_Q$, $V_g$ and $V_{\mathit{NR}}$ are defined in Fig.~\ref{fig:fig1}, and $e$ is the magnitude of the electron charge. The CPB is often in DC-SQUID configuration (Fig.~\ref{fig:fig1}), and thus $E_J(\Phi)$ represents the effective Josephson energy of the SQUID, which can be tuned in situ by adjusting an applied magnetic flux $\Phi$. Importantly, the relative magnitude of the electrostatic and Josephson terms determines the nature of the CPB's energy eigenstates. For example, if $E_J/4E_C\ll 1$ then the energy eigenstates are essentially charge states $\ket{n}$ (i.e. eigenstates of the Cooper-pair number operator $\hat{n}$), except at  charge degeneracy points, which are defined by $n_{\Sigma}\approx (2n+1)/2$. At these points adjacent charge states are mixed and the system is well characterized by the two-state charge qubit model.\cite{Makhlin01}  On the other hand, if $E_J/4E_C\gtrsim1$, the energy eigenstates are no longer charge states and instead are composed of weighted superpositions of several $\ket{n}$. And in the limit that $E_J/4E_C\gg1$, the CPB is in the transmon regime.\cite{koch2007charge} 

It is presumed that the nanoresonator can be modeled via Eq. (\ref{HamiltonianNR}) as a quantum simple harmonic oscillator in the usual manner: $\hat{a}^{\dagger}$ and $\hat{a}$ are creation annihilation operators for the mechanical mode, which is generally assumed - but not necessarily limited - to be one of the fundamental flexural modes of the suspended nanostructure (the fundamental in-plane mode for the device in Fig.~\ref{fig:fig1}; the fundamental out-of-plane mode for the membrane resonator in Ref. \citen{pirkkalainen2013hybrid}); $\omega_{\mathit{NR}}/2\pi$ is the mode's frequency; and $\hbar$ is Planck's constant.  Finally, Eq. (\ref{HamiltonianINT}) represents the electrostatic coupling that is established between the motion of the nanoresonator and the charge on the CPB island. Here, the scale of the coupling strength is set by the prefactor $\lambda$, which is given by\cite{lahaye2009nanomechanical} 
\begin{equation}
\lambda=-4\frac{E_C}{\hbar}\frac{\mathrm{d}C_{\mathit{NR}}}{\mathrm{d}x}\frac{V_{\mathit{NR}}}{e}x_{zp},
\label{couplingNR}
\end{equation}
where $x_{zp}=\sqrt{\hbar/2m\omega_{\mathit{NR}}}$ represents the zero-point motion of the mechanical mode and $m$ is its effective mass, defined by $m=\alpha\rho wLt$, where $\alpha=\int_{-L/2}^{L/2}U(z)^2\mathrm{d}z$, $\rho$ is the mass density of the nanostructure, and the geometrical dimensions $w$, $t$, and $L$ are defined in Fig.~\ref{fig:fig1}(b).  The quantity $U(z)$ is the displacement of the neutral axis\cite{cleland2002foundations} as a function of position $z$ along the beam  [Fig.~\ref{fig:fig1}(b)].  It is important to note that the value of $\alpha$, and hence $m$, will depend upon the choice for normalization of $U(z)$ - e.g. whether $U(z)$ is normalized so that the displacement $x_{zp}$ represents the zero-point motion of the nanostructure's center of mass, or the average zero-point motion of the structure over the length of the CPB electrode $L_{e}$, or any other arbitrary convention.  However, because $\tfrac{\mathrm{d}C_{\mathit{NR}}}{\mathrm{d}x}\propto\int_{-L_{e}/2}^{L_{e}/2}U(z)\mathrm{d}z$, $\lambda$ itself is independent of the definition of $x$, as one should expect.      
 
Experiments to date provide strong evidence that Eqs. (\ref{hamiltonian}) to (\ref{couplingNR}) give an accurate accounting of the dynamics of capacitively-coupled CPBs and nanomechanical resonators in a semi-classical limit where the mechanical mode is driven to a large amplitude with effective number state populations of $\sim10^3$ to $10^6$.\cite{lahaye2009nanomechanical,suh2010parametric,pirkkalainen2013hybrid}  However, experiments fully in the quantum regime, where many of the proposals in the literature could be implemented, remain to be achieved.    The primary roadblocks are technical in nature and derive from having to simultaneously satisfy the following conflicting demands:  establishing strong coupling $\lambda$ between the qubit and the nanoresonator; maintaining long CPB coherence times, which from here on will be denoted generically by $T_{2}$ or, when appropriate, the inhomgeneously broadened coherence time $T_2^*$; and achieving low thermal occupation numbers $N_{th}$ in the mechanical mode. In the following section we discuss these interconnected criteria in greater detail. 

\subsection{Challenges in the Development of Coupled CPB-Nanoresonator Systems}
It has been recognized for more than a decade that CPB-coupled nanoresonators can serve as testbeds for studying quantum properties of mechanical systems.  However, experiments have yet to catch up with the theoretical ideas in this field.  The main challenge has been engineering strong coupling between the two systems while simultaneously minimizing the interactions of the individual systems with the environment.  Generally speaking, this requires establishing CPB-nanoresonator coupling strengths that exceed the decoherence rates of the nanoresonator and CPB,  $\kappa$ and $\gamma$ respectively. Heuristically, what this \textit{strong coupling} requirement implies is that the two systems exchange energy or information with each other at a faster rate than with unaccounted for degrees of freedom.  

In the two experiments with CPB-coupled nanoresonators thus reported, coupling strengths $\lambda/2\pi> 1\,{\rm MHz}$ were achieved,\cite{lahaye2009nanomechanical,suh2010parametric,pirkkalainen2013hybrid} which exceed some of the best reported CPB decoherence rates ($\gamma/2\pi = 0.7\,{\rm MHz}$ for a charge qubit embedded in a CPW cavity\cite{wallraff2004strong} and $\gamma/2\pi=10\,{\rm kHz}$ for a single-junction transmon in a 3D waveguide\cite{rigetti2012superconducting}). Moreover, such coupling strengths are larger than typical linewidths of flexural nanoresonators at milli-Kelvin temperatures ($ \sim 1\,{\rm kHz}$ for Ref. \citen{lahaye2009nanomechanical} and $\sim 10\,{\rm kHz}$ for Ref. \citen{pirkkalainen2013hybrid}), which should set the scale of $\kappa$ in the quantum regime.  However, in both cases  the mechanical resonators were greatly detuned in energy from the CPBs ($\omega_{\mathit{NR}}/2\pi = 60 - 70\,{\rm MHz}$ versus $\Delta E_{\mathit{CPB}}/h \sim 4 - 10\,{\rm GHz}$), thus precluding the study or use of coherent, resonant interactions between the systems.\footnote{It should be noted that in Ref. \citen{pirkkalainen2013hybrid} transitions between electromechanical dressed-states were observed, however, this was accomplished by driving the mechanical element into an essentially classical regime with $\geq 10^3$ quanta in the mechanical mode.}  In this far-detuned (dispersive) limit, a more appropriate figure for comparison is really the dispersive coupling strength, given by\cite{irish2003quantum,lahaye2009nanomechanical} 
\begin{equation}
\frac{\chi}{2\pi}=\frac{\hbar\lambda^2E^2_J}{\pi\Delta E_{\mathit{CPB}}(\Delta E^2_{\mathit{CPB}}-(\hbar\omega_{\mathit{NR}})^2)},
\label{dispersive}
\end{equation} 
which would set the time scale for generating Schr\"{o}dinger cat states of the mechanics\cite{armour2008probing} and  limits CPB transition linewidths for performing number state detection\cite{clerk2007using} using dispersive techniques.  For both experiments to date, $\chi/2\pi \sim 1\,{\rm kHz}$, comparable to the nanoresonators' linewidths, but orders of magnitude less than the decoherence rates of the qubits used for those experiments (and at least an order of magnitude less than the best $\gamma$ demonstrated thus far in cQED), making such quantum measurement infeasible with these first devices.      

On the face of it, there would appear to be multiple, independent paths toward further development of CPB-coupled nanoresonators for advanced quantum measurement:  improve CPB-nanoresonator coupling strengths; engineer long CPB coherence times; and increase the nanoresonator's frequency. However, these three directions are interdependent, and modifications to enhance one parameter may, in some cases, adversely impact another. For example,  Eq. (\ref{couplingNR}) suggests that CPB-nanoresonator coupling can be maximized by working with as large a charging energy $E_C$ as possible.  This makes sense: the larger $E_C$ is, the greater the charge dispersion (or sensitivity to changes in polarization charge $n_g$) and hence the more strongly one can couple the motion of a nearby suspended electrode.  But, unfortunately, increasing $E_C$, for the same reasons, also increases the CPB's suceptibility to local charge noise, whether it arises from trapped surface charge fluctuators, two-level systems (TLS) or non-equilibrium quasiparticle tunneling.\cite{schuster2007circuit} This yields short  coherence times (typically $T_2\ll 1$ $\mu$s), as well as slow drifts  and random jumps in the system's energy that makes these devices notoriously difficult to work with.   For this reason, the superconducting quantum computing community has abandoned CPBs in the charge qubit regime and moved to low-$E_C$ transmons, which, as noted above have given the longest coherence times to date for any superconducting qubit, approaching 100 $\mu s$.\cite{rigetti2012superconducting} Thus, moving to the transmon regime would also appear to be the right direction for mechanics, as was done in Ref. \citen{pirkkalainen2013hybrid}. However, it is crucial to point out that the resulting 30-to-40-fold reduction in $E_C$ (in moving from typical charge qubit values to typical transmon values), without making any additional changes, leads to a reduction in $\chi$ of $\sim 1000$, essentially leaving the product of $\chi T_2$ unchanged at best. Additional solutions are thus required to reach the strong coupling regime.

Moving forward, coupling strength can still be improved by several means while working in the transmon regime:\footnote{It should also be noted that in Ref. \citen{pirkkalainen2013hybrid} the factor of 40 reduction in $E_C$, in comparison with Ref. \citen{lahaye2009nanomechanical}, was made up for by a $\sim$ 1000-fold increase in $\mathrm{d}C_{\mathit{NR}}/\mathrm{d}x$ by utilizing a plate-style geometry.  Taking into account the much larger mass of the plate, this yielded a maximum coupling of $\lambda/2\pi= 4.5\,{\rm MHz}$, a factor of two larger than in Ref. \citen{lahaye2009nanomechanical}, but achieved using one-third the value of $V_{\mathit{NR}}$.  Unfortunately, this was not enough to achieve strong coupling, due in part to the short coherence time of the transmon, which was observed to be $T_2^*\sim 70\,{\rm ns}$ and thought to be limited by quasiparticle poisoning.} increasing the applied voltage $V_{\mathit{NR}}$; increasing $\mathrm{d}C_{\mathit{NR}}/\mathrm{d}x$, and utilizing low-frequency, high-aspect-ratio devices  (i.e. $L/w\gg1$) .  Increasing the voltage would appear to be a simple approach. However, it is not yet clear whether doing so leads to a degradation of $T_2$ due to the increased charge noise that arises from the application of $V_{\mathit{NR}}$ and the large electric fields ($\sim 10^6\,{\rm V}/{\rm cm}$) in between the nanoresonator electrode and nearby electrodes like the CPB island. Tailoring geometries and materials to engineer larger $\mathrm{d}C_{\mathit{NR}}/\mathrm{d}x$ would also seem to be a straightfoward approach.  Nonetheless, it too is not trivial.  If special precautions are not taken to limit the bandwidth of the external bias circuitry, the CPB (transmon or not) will experience radiative damping with a rate given by $\Gamma =\Delta E^2_{\mathit{CPB}}C^2_{\mathit{NR}}Z_0/\hbar^2C_{\mathit{CPB}}$,\cite{houck2008controlling} where $C_{\mathit{CPB}}$ is the CPB's total effective island capacitance and $Z_0$ is the impedance of the external bias circuitry.  The resulting relaxation time $T_1$ can be quite low; for example, $C_{\mathit{NR}} = 5\,{\rm fF}$, $C_{\mathit{CPB}}=50\,{\rm fF}$, $\Delta E_{\mathit{CPB}}/h=5\,{\rm GHz}$, yields $T_1=1/\Gamma = 40\,{\rm ns}$ and a maximum coherence time of $T_2=2T_1=80\,{\rm ns}$.  Thus proper engineering of the bias circuitry's impedance and bandwidth is also critical for maintaining CPB coherence times.  Both of these effects are currently being researched by the authors (see Section \ref{subsection:sub3}), who, in unpublished work, have seen in spectroscopic measurements that transition linewidths  ($\sim1/T_2^*$) as narrow as $2\,{\rm MHz}$ persist in a voltage-biased transmon up to at least $V_{\mathit{NR}}= 8\,{\rm V}$, where superconducting band-stop filters\cite{hao2014development} are used to limit the radiative decay of the bias channel.        
 
Through simple considerations, one can show from Eq. (\ref{couplingNR}) that $\lambda$ scales as $L^{3/2}/w$, motivating the use of high-aspect-ratio nanostructures to reach the strong coupling regime.  Of course, because flexural mode frequencies scale as $w/L^2$,\footnote{For example, the in-plane flexural mode frequencies of a thin beam, considering pure bending, are given by $\omega_{\mathit{NR}}/2\pi=\frac{a^2_iw}{L^2}\sqrt{\frac{E}{12\rho}}$,\cite{cleland2002foundations} where $E$ is the Young's modulus of the material and $a_i = 4.73, 7.89, 10.99$ for the first, second and third modes respectively.}  taking this approach would lead to greatly reduced mode frequencies. For instance, a factor of 10 increase in $\lambda$ compared with Ref. \citen{lahaye2009nanomechanical}, achieved by increasing only the length, would require extending $L$ by a factor $10^{2/3}\sim4.6$.  This would result in $\omega_{\mathit{NR}}/2\pi \sim 3\,{\rm MHz}$, which would have a large thermal population even at milli-Kelvin temperatures (e.g. $N_{TH} \sim 140$ at $T= 20\,{\rm mK}$).  Side-band cooling techniques developed in cavity mechanics\cite{teufel2011sideband} could be utilized for ground-state cooling of a mechanical mode prior to coupling to the CPB. However, it is expected that the thermal relaxation rate of the mode would be greatly increased, proportional to $\kappa N_{TH}$, which in turn would place more stringent constraints on nanoresonator Q-factors for achieving the strong coupling regime. Moreover, an additional concern for increasing the aspect-ratio is a resulting decrease in the voltage at which ``pull-in" occurs\cite{buks2001metastability} - this is the voltage at which the nanostructure becomes unstable and snaps into the bias electrode, which usually leads to stiction between the mechanical element and the electrode.  As a rule of thumb, the pull-in voltage goes as $V_{Sn}\approx\sqrt{8kd^2/27C_{\mathit{NR}}}$, where $k$ is the effective spring constant and $d$ is the zero-voltage spatial gap between the structure and the electrode.  Because $k\propto w^2/L^3$ and $C_{\mathit{NR}} \propto L$, one sees that $V_{Sn} \propto w/L^2$, which could be on the order of volts or much less for high-aspect-ratio devices. The gain in coupling strength by increasing the device's length can thus be completely cancelled out (and actually reversed) by the reduction in pull-in voltage. Pull-in voltage could also be problematic for achieving large $\lambda$ using graphene or carbon nanotube nanoresonators, due their greatly reduced $k$.    

Of course, the considerations in the previous paragraph are rather imprecise, and detailed modeling using finite element simulations and incorporating Casimir forces\cite{buks2001metastability} would be necessary to find optimal sets of parameters for maximizing $\lambda$ over different configurations.  Nonetheless, the expressions for $V_{Sn}$ and $\lambda$ gives rise to the following rule of thumb for maximum coupling achievable for fundamental flexural modes:   
\begin{equation}  
\lambda_{max}\approx-\frac{8E_{c}}{\hbar}\sqrt{\beta\frac{\hbar\omega_{\mathit{NR}}C_{\mathit{NR}}}{27e^2}},
\label{couplingmax}
\end{equation}
which arises from substituting $V_{Sn}$ in to Eq. (\ref{couplingNR}). Here $\beta$ is a constant of order unity that accounts for the deviation of $\mathrm{d}C_{\mathit{NR}}/\mathrm{d}x$ from a parallel-plate approximation.  Using typical parameters for transmon-type CPBs and UHF flexural resonators, Eq. (\ref{couplingmax}) suggests that $\lambda_{max}$ could approach 100 MHz, if coupling voltages approaching $V_{\mathit{NR}} \sim 30\,{\rm V}$ can be applied.  It remains to be seen whether this can be achieved with CPB-coupled nanoresonators.


Finally, it should be noted that one additional possibility for increasing the dispersive coupling $\chi$ without increasing $\lambda$ is to decrease the detuning in energy between the nanoresonator and qubit.   There are clearly two ways to do this:  increase nanoresonator frequencies; or decrease qubit transition energies $\Delta E_{\mathit{CPB}}$.  For the former case, it should be possible to engineer flexural nanoresonators with third mode frequencies in the range of 3 GHz; transmon type qubits could then be tuned via magnetic flux close to resonance with these mechanical modes in the same way as is done in cQED\cite{wallraff2004strong} with cavity resonators. This technique is currently being investigated by the authors using a device similar to one shown in Section \ref{subsection:sub3}.  For the latter case, fluxonium\cite{pop2014coherent} devices could be substitued for the transmons. Fluxonium, while relatively new in the lineage of superconducting qubits, has demonstrated long coherence times ($T_2^*= 14\,\mu{\rm s}$) at transition energies as low as $\sim 500\,{\rm MHz}$,\cite{pop2014coherent} which would be nearly resonant with the fundamental modes of properly engineered nanobeams.  This approach remains the subject of future work.   
\section{Experimental Development of CPB-based Nanoresonator Read-Out at Syracuse} \label{sec:sections}
In this section we highlight some of the experimental efforts  in recent years at Syracuse to develop QEMS that integrate multiple superconducting devices and circuitry with nanomechanical systems.  In particular, these new hybrid devices are composed of the three main components: superconducting CPW cavities; CPB-based qubits; and suspended superconducting wires as flexural nanomechanical elements.  Sample characteristics and data for three generations of devices, extending back to 2012, are discussed.  It our intent that this section serve not only as a record of what has been accomplished, but also to provide greater context to the challenges discussed in the previous section and to serve as a guide for those who might soon be interested in taking up the challenge.  
 
\subsection{Generation I: CPB in the Charge Regime Coupled to a Lumped-Element LC Circuit and Flexural Nanoresonator} 
\begin{figure}
\begin{center}\includegraphics[ width=.95\columnwidth,keepaspectratio]{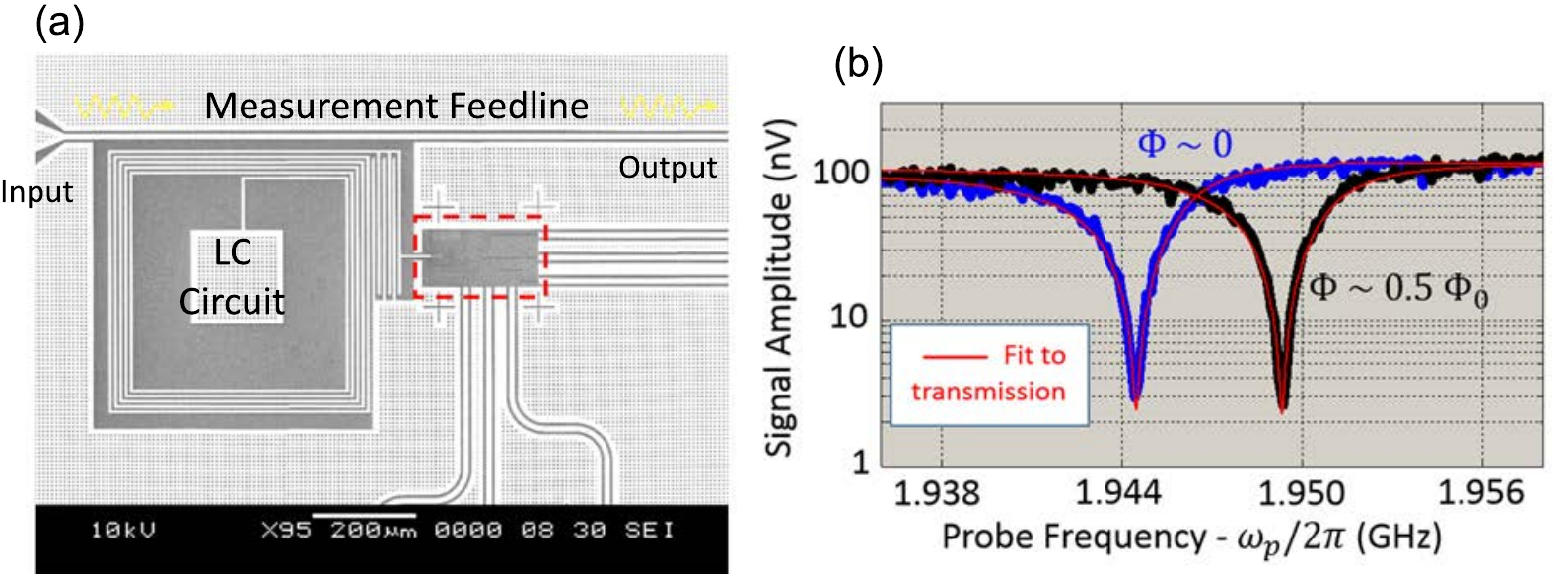}\end{center}
	\caption{(a) SEM micrograph displaying a birds-eye-view of the detection circuit for measuring the first generation of CPB-type QCNR. The detection circuit consists of a lumped element inductor and capacitor that are capacitively coupled to the CPB (not visible in this image, but located in the region denoted by the red, dashed rectangle). The frequency of the LC circuit serves as a proxy for CPB state through a simple dispersive interaction, which thus enables measurements of the CPB absorption spectrum.  The frequency response of the LC is probed by performing transmission measurements of the measurement feedline which is both capacitively and inductively coupled to the LC circuit. (b) Transmission measurements of the feedline for two different values of $\Phi$ applied to the CPB.}
	\label{fig:fig2}
\end{figure}

In 2012 the LaHaye group began fabrication and measurement of its first generation of CPB-coupled nanoresonators (Fig.~\ref{fig:fig2}), with the goal of utilizing the CPB to perform dispersive, number-state read-out of UHF-range nanomechanical elements at low thermal occupation numbers. This was to be accomplished by performing measurements of the CPB's absorption spectrum to look for the mechanical Stark shifts in the CPB's transition energy that should arise as a result of the dispersive interaction with the nanoresonator.\cite{clerk2007using}  In the following paragraphs, some of the key design considerations for Generation I are discussed. 

In an initial attempt to balance the conflicting dependence on $E_C$ of coupling strength and dephasing due to charge noise, the CPB parameters were chosen so that the device resided in between the charge qubit and transmon regimes. Specifically, the chosen geometry yielded $E_{J0}/E_{C}\approx 6$ and $E_{C}/h \approx 1.8\,{\rm GHz}$; the value of $E_C$ is consistent - to within design tolerances - with electrostatics simulations of the geometry using ANSYS Q3D, which yield $E_C/h=2\,{\rm GHz}$.  

The CPB was embedded within a planar, lumped-element LC circuit [Fig.~\ref{fig:fig2}(a)], which was to serve the purpose of both filtering the CPB's electromagnetic environment and also providing a means for performing spectroscopy of the CPB to measure its absorption spectrum. The coupling between the two systems was provided by an inter-digitated capacitor $C_Q=5\,{\rm fF}$ as calculated using Q3D.  In contrast to typical applications in cQED where CPW or 3D cavities are used for isolation and read-out, the effective L and C were chosen to yield a low resonance frequency $\omega_{LC}$, in the range of 1 to 2 GHz.  The chosen geometry resulted in $\omega_{LC}/2\pi=1.94 -1.95\,{\rm GHz}$ [Fig.~\ref{fig:fig2}(b)], which was in good agreement with Sonnet simulations that predicted 1.93 GHz.  The LC was engineered to be over-coupled to a measurement feedline in order to provide fast and efficient measurement.\cite{johansson2006fast} Fits to the feedline response\cite{megrant2012planar} [Fig.~\ref{fig:fig2}(b)] determined a loaded quality factor of $Q_L = 300 - 500$ and intrinsic quality factor $Q_i=12-15 \times 10^3$, depending on flux $\Phi$ applied to the CPB.  The initial purpose of using a low-frequency, lumped-element LC resonance was to insure that the circuit would be far-detuned in energy from the CPB and thus interacting in the weak dispersive regime, where dephasing effects and modifications of the CPB's absorption spectrum would be minimal. However, because of initial miscalculations which led to a larger than desired $C_Q$, the two systems interacted very strongly as discussed below.  As well, it was thought that introducing the DC voltage bias $V_{\mathit{NR}}$ would be technically less challenging with the lumped-element design than with a distributed resonator.\cite{chen2011introduction}

\begin{figure}
\begin{center}\includegraphics[ width=.95\columnwidth,keepaspectratio]{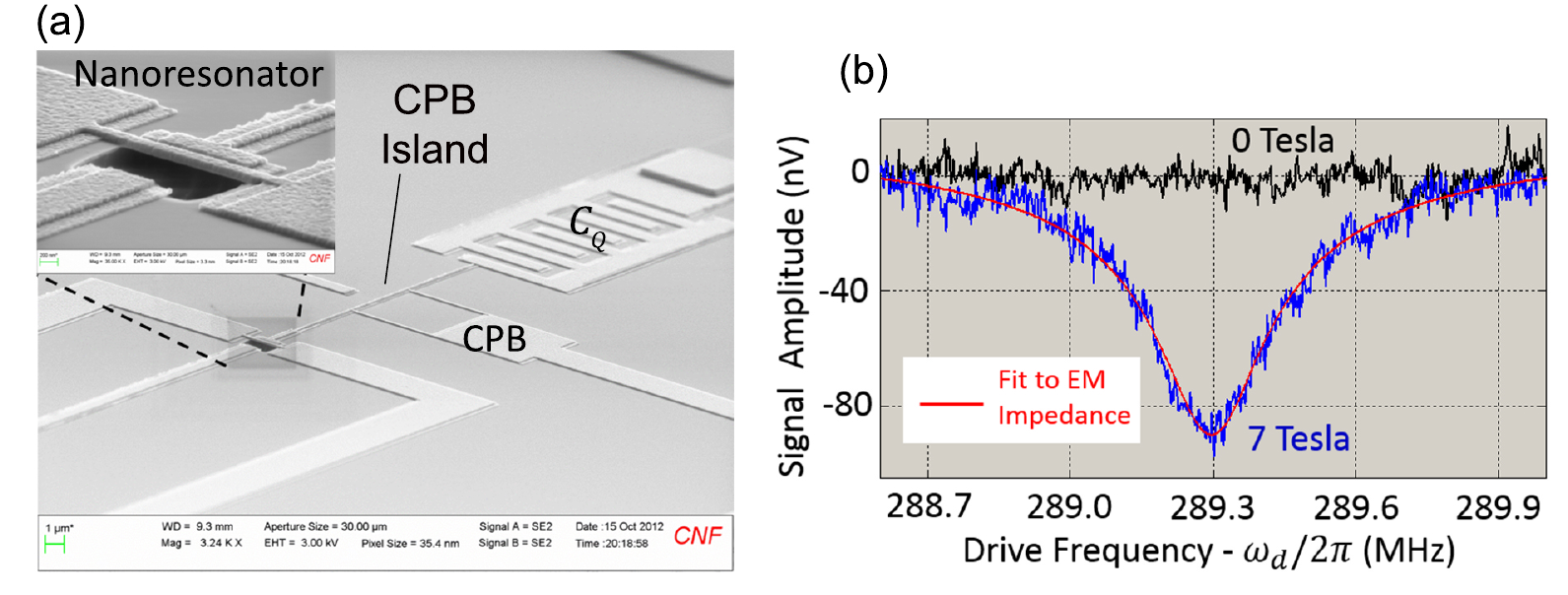}\end{center}
	\caption{(a) SEM micrograph displaying the region in Fig.~\ref{fig:fig2} denoted by the red, dashed rectangle.  This region includes the CPB and nanostructure from the first generation of QCNR developed and measured by the authors. The inset shows a close up of the aluminum nanostructure.  The fundamental in-plane flexural mode of this structure should couple most strongly to the charge on the CPB island. From COMSOL simulations and analytical calculations, this mode should have a resonant frequency of $\omega_{\mathit{NR}}/2\pi \approx 300\,{\rm MHz}$. (b) Magnetomotive measurements of the fundamental mode response at $T=4\,{\rm K}$ are in good agreement with the expected frequency from simulations.}
	\label{fig:fig3}
\end{figure}

The nanostructure was fabricated out of aluminum using standard plasma etching. The geometric parameters of the structure, $w=200\,{\rm nm}$, $t=100\,{\rm nm}$, $L=1.8\,\mu{\rm m}$, $d=70\,{\rm nm}$ [Fig.~\ref{fig:fig3}(a)], were chosen to give a fundamental in-plane flexural resonance frequency of $\omega_{\mathit{NR}}/2\pi=300\,{\rm MHz}$ and coupling capacitance $C_{\mathit{NR}}=180\,{\rm aF}$ as calculated using finite element simulations.  Measurements of the resonator's frequency at $T=4\,{\rm K}$ using magnetomotive detection\cite{cleland1996fabrication} [Fig.~\ref{fig:fig3}(b)] were in good agreement with the simulations of the mechanics. From these parameters, estimates of the maximum CPB-nanoresonator coupling using Eq. (\ref{couplingmax}) yielded $\lambda_{max}/2\pi=50-100\,{\rm MHz}$, depending on the value of $\beta$, which, from simulations, should have been on the order of 0.2 or larger. For such values of coupling strength, the dispersive interaction should have reached $\chi/2\pi> 1\,{\rm MHz}$. Based upon estimates from Ref. \citen{clerk2007using}, this would have been sufficient for the number-state statistics of the nanoresonator to be resolvable, even for a thermal state of the nanoresonator at $T=30\,{\rm mK}$ ($N_{TH}\approx2$), provided that the decoherence rate of the CPB satisfied $\gamma \lesssim 1\,{\rm MHz}$, which has been observed previously for CPBs in cQED architectures.\cite{wallraff2004strong} Moreover, the quality factor $Q_{\mathit{NR}} \approx 1000$ of the nanoresonator measured at $T=4\,{\rm K}$ using magnetomotive detection strongly suggested that the nanoresonator decoherence rate would satisfy $\kappa/2\pi< 1\,{\rm MHz}$ at milli-Kelvin temperatures as well.     

Samples were  mounted in a light-tight copper box that was anchored to the mixing chamber (MC) of a dilution refrigerator and cooled down to $T\lesssim 30\,{\rm mK}$.  Microwave lines for probing the transmission of the measurement feedline were filtered and isolated using standard techniques:  the input feedline had $\sim$ 70 dB of attenuation inside the refrigerator, with cryogenic attenuators rigidly anchored to the 1K, still, cold-plate and MC stages; and two cryogenic isolators, nominally with 15 dB each of isolation, were located between the output of the feedline and the input of a cryogenic HEMT amplifier anchored to the 4K stage.  DC lines for applying the CPB gate voltage bias $V_g$ and the nanoresonator coupling $V_{\mathit{NR}}$ were heavily filtered using lossy, stainless steel coaxial cables and homemade powder filters at multiple stages, resulting in $>100\,{\rm dB}$ of attenuation for frequencies above $1\,{\rm GHz}$. A homemade superconducting Helmholtz coil bolted to the top of the sample holder was used to provide the magnetic field to control the flux $\Phi$ applied the CPB.  
\begin{figure}
\begin{center}\includegraphics[ width=.95\columnwidth,keepaspectratio]{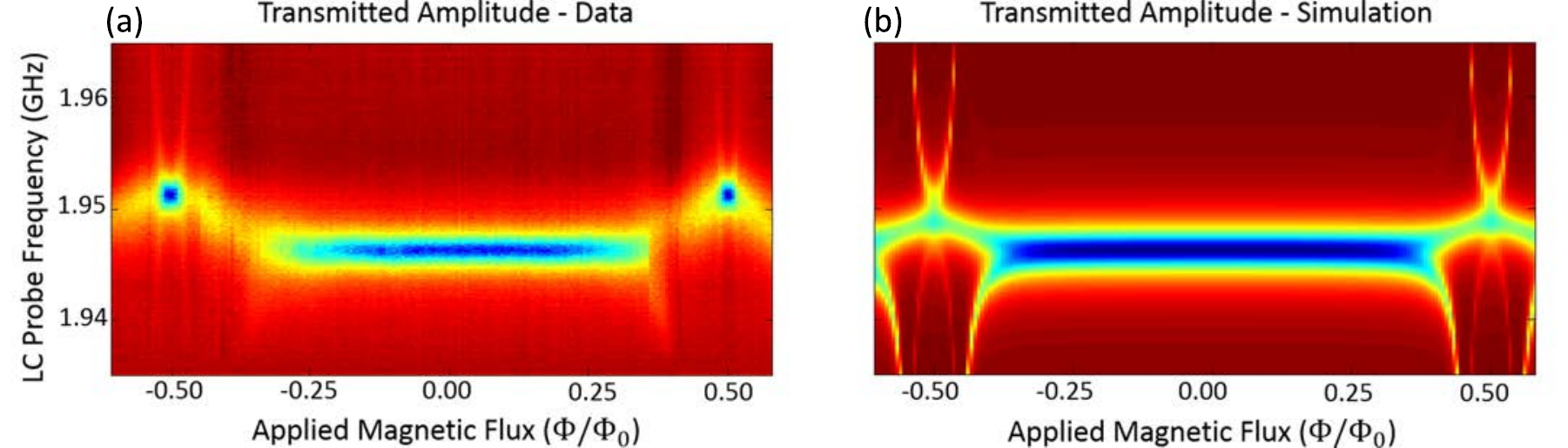}\end{center}
	\caption{Single-tone spectroscopy of the LC circuit and CPB in Figs. \ref{fig:fig1} to \ref{fig:fig3} performed at $T\lesssim 30\,{\rm mK}$ over one flux period, $\Delta\Phi = \Phi_0$. (a) Amplitude response of the LC reveals avoided level crossings that are indicative of the hybridization of the LC and CPB energy levels for values of $\Phi$ where $\Delta E_{\mathit{CPB}}\approx \hbar\omega_{LC}$.  (b) Numerical simulations of the amplitude of the LC circuit's frequency response using linear response theory agree well with the data, reproducing the main features in the spectrum. Simulations were carried out for the following values: $E_{J0}/h=12.7\,{\rm GHz}$, $E_C/h=1.3\,{\rm GHz}$, $\lambda_{LC}/h=160\,{\rm MHz}$, and average LC photon number $\bar{N}=0.3$.}
	\label{fig:fig4}
\end{figure}

Before applying the nanoresonator coupling voltage $V_{\mathit{NR}}$, measurements were conducted to make sure that the LC circuit could be used to read-out the CPB. To a good approximation, the capacitively-coupled LC circuit and CPB can be described in a manner formally analogous to the CPB-coupled nanoresonator, with dynamics captured also by Eqs. \ref{hamiltonian} to \ref{couplingNR}, except with the coupling strength given by 
\begin{equation}
\lambda_{LC}=\frac{4E_{C}C_Q}{e\hbar}\sqrt{\frac{\hbar\omega_{LC}}{2C_T}},
\label{lccoupling}
\end{equation}
where $C_T$ is the total capacitance of the LC circuit.  Using $C_T=340\,{\rm fF}$, as calculated by Q3D, and the previously noted values of $E_C$ and $C_Q$,  the coupling strength was estimated to be quite large: $\lambda_{LC}/2\pi \approx 200\,{\rm MHz}$. As a consequence of the large coupling, Jaynes-Cummings physics could readily be observed in spectroscopic measurements of the coupled LC-CPB systems (Fig.~\ref{fig:fig4}). This was particularly evident in single-tone spectroscopy measurements\cite{schuster2007circuit} where microwaves in the frequency range near $\omega_{LC}$ were applied to the system through the measurement feedline.  By monitoring the amplitude [Fig.~\ref{fig:fig4}(a)] and phase (not shown) of signals transmitted through the feedline using standard heterodyne detection, and varying the flux $\Phi$ applied to the CPB, hybridization of the CPB and LC energy levels could be observed around values of $\Phi$ where $\hbar\omega_{LC}=\Delta E_{\mathit{CPB}}$. The hybridization manifested in the usual avoided level crossings that appear periodically as a function of $\Phi$ with a period of one flux quantum $\Phi_0$, as expected from the dependence of $E_J$ on $\Phi$. Numerical simulations of the transmission measurement versus $\Phi$ and LC probe frequency $\omega$ were performed using linear response theory and the analog of Eqs. (\ref{hamiltonian}) to (\ref{HamiltonianINT}) with Eq. (\ref{lccoupling}) for the CPB-coupled LC. The simulations [Fig.~\ref{fig:fig4}(b)] incorporated 50-50 averaging of $n_g=0$ and $n_g=0.5$ to account for quasiparticle poisoning,\cite{riste2013millisecond} which is believed to have been occurring on a much faster time scale than the  measurement time at each value of $\Phi$. These results were seen to agree well with measurements, capturing many of the features seen in the spectroscopy. 
\begin{figure}
\begin{center}\includegraphics[ width=.95\columnwidth,keepaspectratio]{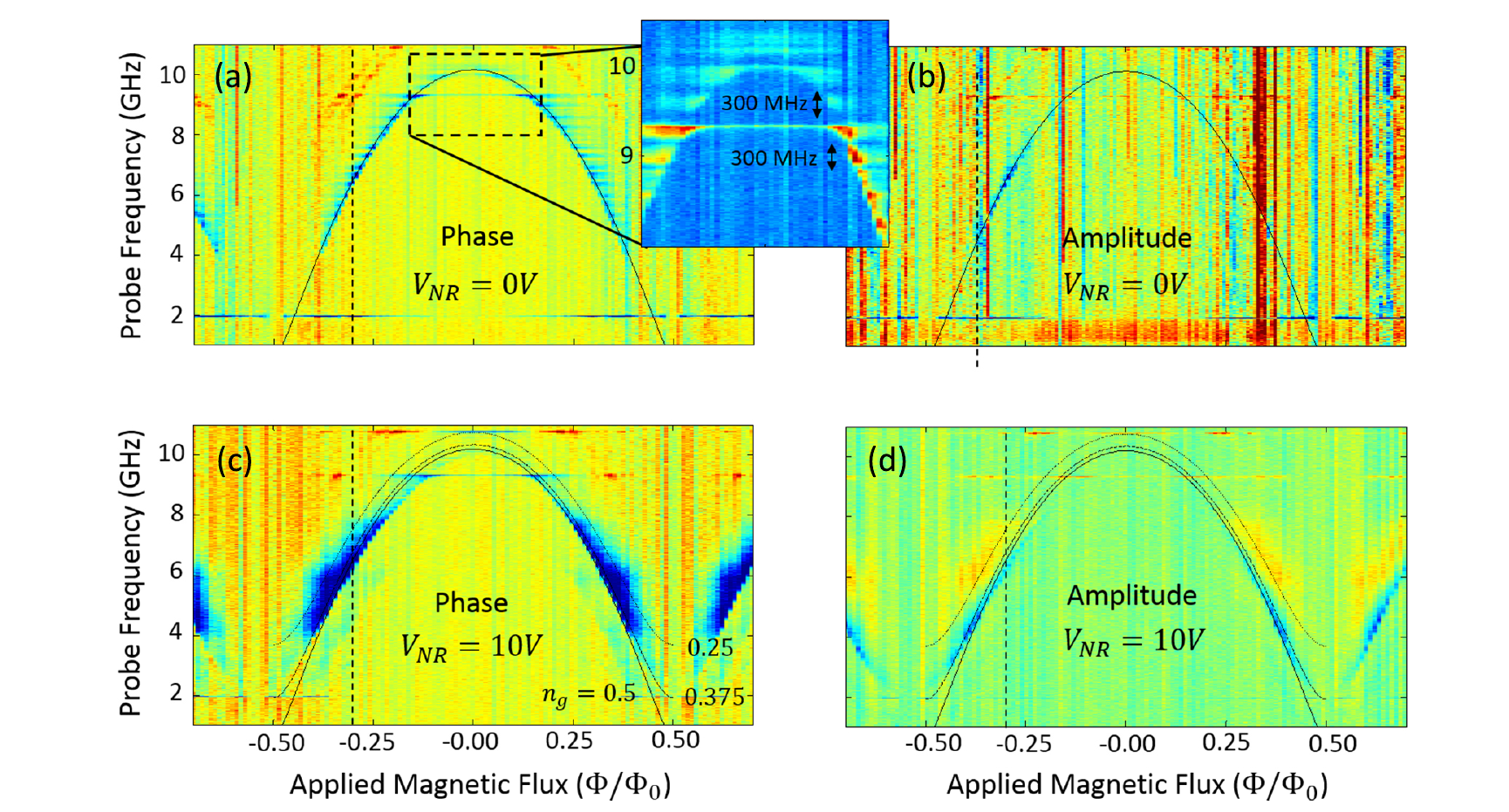}\end{center}
	\caption{Two-tone spectroscopy maps of the first generation of QCNR developed at Syracuse shown in Figs. \ref{fig:fig1} to \ref{fig:fig3} versus $\Phi$ at $T\lesssim 30\,{\rm mK}$. (a) and (b) Phase and amplitude of the LC circuit's response for $V_{\mathit{NR}}=0\,{\rm V}$. The inset shows a higher resolution spectroscopy scan of the top of the hyperbole to illustrate the regular spacing of the avoided level crossings. The color scale in the inset has been reversed to enhance viewing contrast while superimposed on top of the main figures.  (c) and (d) Phase and amplitude of the LC circuit's response for $V_{\mathit{NR}}=10\,{\rm V}$. The black hyperbole in (a)-(d) indicate the lowest-order transition energy of the CPB, $\Delta E_{\mathit{CPB}}$, versus $\Phi$ and were generated from numerical calculations using Eq. (\ref{HamiltonianCPB}) and the following parameters: $E_C =1.8\,{\rm GHz}$, $E_{J0}=11.7\,{\rm GHz}$, and $n_g=0.5.$  Also plotted in (c) and (d) are values of $\Delta E_{\mathit{CPB}}$ for $n_g=0.25$ and $n_g=0.375$, which are denoted by dotted and dashed lines respectively. The dashed vertical lines indicate locations of the individual traces shown in Fig.~\ref{fig:fig6}.}
	\label{fig:fig5}
\end{figure}

Two-tone, continuous-wave spectroscopy\cite{schuster2007circuit} of the CPB and LC was perfomed next in order to measure the absorption spectrum of the CPB over the full-range of $\Delta E_{\mathit{CPB}}$ - the CPB's lowest transition energy - as a function of $\Phi$  (Figs. 5 and 6).  These measurements were conducted by first fixing $\Phi$, and then applying two microwave tones to the CPB-coupled nanoresonator. The first tone, $\omega$, was applied to the LC circuit and fixed at $\omega=\omega_{LC}$ to probe the LC circuit's response to changes in the CPB's state; the second, spectroscopy tone $\omega_s$, was then applied to excite Rabi oscillations in the CPB.  The average amplitude and phase of the signal transmitted at $\omega$ was then recovered using heterodyne detection.  Measurements were typically repeated over a large range of  $\omega_s$, from 0.5 GHz to 11 GHz, and one flux period $\Phi_0$. Results from two sets of measurements are shown in Fig. 5.  It is clear that the envelope of the CPB's absorption spectrum is in good agreement with the predicted lowest-energy transition $\Delta E_{\mathit{CPB}}$ (solid hyberbolic line), which was calculated numerically using Eq. (\ref{HamiltonianCPB}).  However, it also clear that there are many additional features in the spectrum. In fact, higher resolution spectroscopy of the LC circuit phase [inset, Fig. 5(a)] reveals that the main absorption line is broken by a series of approximately regularly-spaced avoided level crossings.  Curiously, in many locations in the spectroscopy map, the spacings in energy between avoided level crossings are $\sim 300$ MHz, comparable to the nanoresonator's energy. Moreover, the location and spacing of the crossings did not appear to depend on power of the spectroscopic tone or probe tone; increasing power simply broadened the features. This suggests that these features were not related to coupling with the LC resonance or any higher-order modes in the extended LC circuit.\footnote{It should be noted that one higher-order mode of the inductor is thought to be seen in the spectrum at $\sim 9.3\,{\rm GHz}$. This would correspond to the third quarter-wave mode of the extended planar inductor; the inductor was shorted to the ground plane at one end.}.  Because the avoided level crossings were observed even with $V_{\mathit{NR}}=0$, it is thought unlikely that these features were due to the nanoresonator.  However, it is possible that an intrinsic DC bias existed between the nanoresonator electrode and the CPB island, which was electrically isolated from all other portions of the circuit, providing the coupling. Such offsets have been reported anecdotally in the literature before,\cite{hertzberg2009back} but in this case no measurements could be performed to confirm whether an offset was present or not.    

\begin{figure}
\begin{center}
\begin{tabular}{c}
\includegraphics[ width=.65\columnwidth,keepaspectratio]{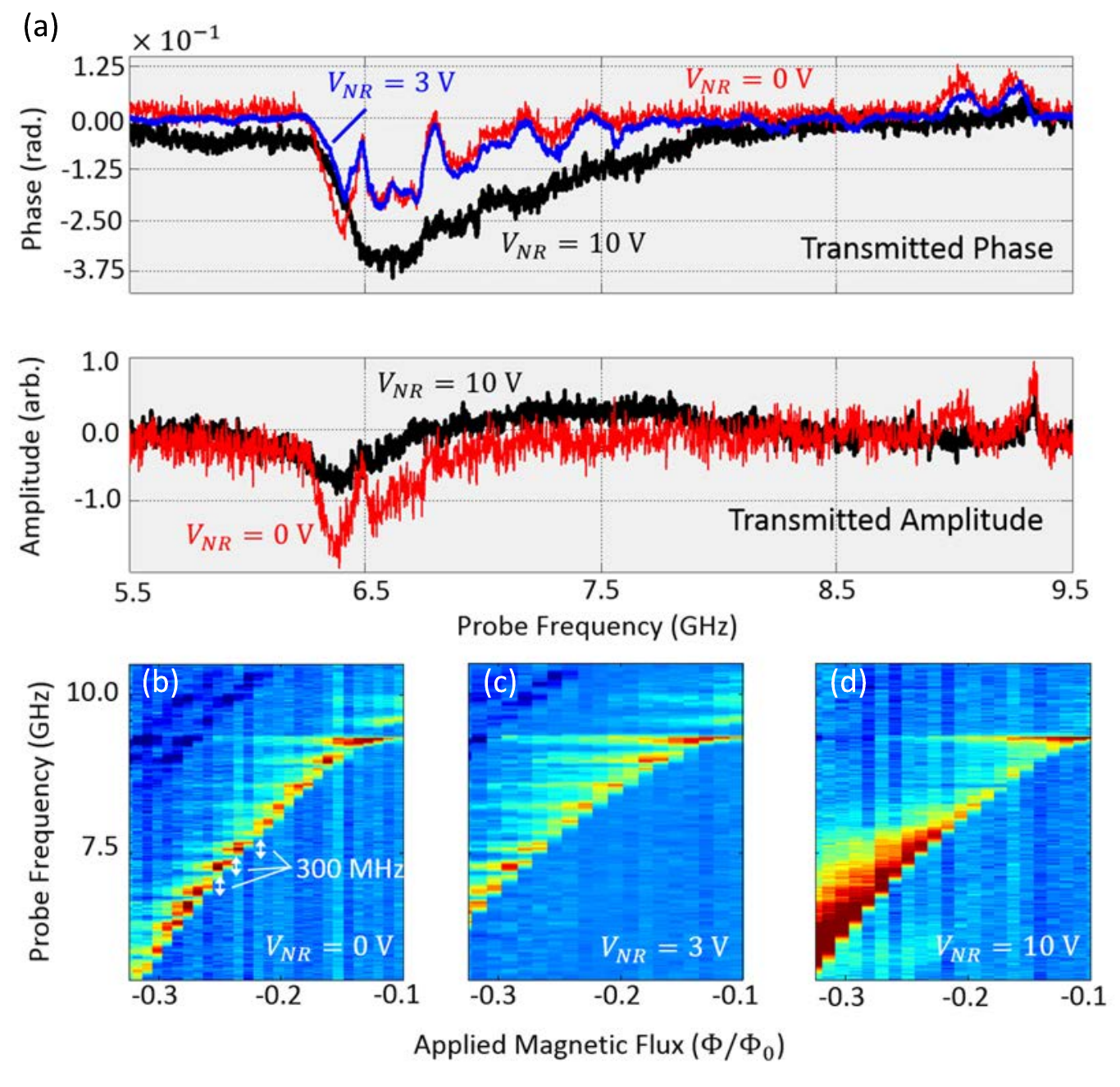}
\end{tabular}	
\end{center}
\caption{Comparison of spectroscopy data for different $V_{\mathit{NR}}$.  (a) Amplitude and phase of individual spectroscopy traces from the maps in Fig.~\ref{fig:fig5} at values of flux denoted by the vertical dashed lines. As $V_{\mathit{NR}}$ is increased, the spacing between the apparent avoided level does not change, but the resonances broaden and ultimately overlap at $V_{\mathit{NR}}=10\,{\rm V}$. This is readily apparent in the spectroscopy ``close-ups" shown in (b), (c), and (d) for 0 V, 3 V, and 10 V respectively. Note that the color scale is reversed in (b), (c) and (d) to enhance the contrast of the resonances.}
	\label{fig:fig6}
\end{figure}

The next step in the measurement process was to increase $V_{\mathit{NR}}$ to probe any changes in the absorption spectrum of the CPB [Figs. 5(c,d) and 6] that resulted from the expected dispersive interaction with the nanoresonator.  Voltages ranging from 0.5 V to 15 V were applied between the nanoresonator and CPB island using a home-made battery-powered source. A motor with a high gear-ratio was used to slowly increment the voltage to the desired value at a rate of mV/sec. This was implemented in order to make the change in charge on CPB electrode as adiabatic as possible and to avoid stirring up excessive charge noise in substrate or on surfaces in the vicinity of the CPB.  When the desired value of $V_{\mathit{NR}}$ was achieved, the motor was powered-down and disconnected from the apparatus. Complete spectroscopy data was taken only up to $V_{\mathit{NR}}=10\,{\rm V}$ (Fig. 5) as the device was destroyed at $V_{\mathit{NR}}=15\,{\rm V}$ when the connection supplying $V_{\mathit{NR}}$ was erroneously removed. For measurements up to 10 V, the locations of the avoided level crossings did not appear to change location in energy nor did the spacing between the features change (Fig. 6).  However, the features became progressively blurred out; this is clearest in Figs. 6(b) to 6(d). Interestingly, the spectroscopy maps at 10 V indicated that the change in amplitude of the probe signal flipped sign when the spectroscopic tone passed through the main CPB absorption line [Figs. 5(d) and 6(a)]. 

Because the first sample was destroyed, the origins of the additional structure in the phase and amplitude of the spectroscopy maps was not determined.  One possibility was that this structure was due to coupling to an array of TLS, as has been previously reported in the literature.\cite{grabovskij2012strain} The exact spectrum of such TLS should be unique from sample to sample, so measurements of a second, identically-designed sample could be used to rule out whether TLS were responsible for the observed splittings.  Thus a second, nominally-identical device  was cooled down. However, the device was defective and spectroscopic signatures of the CPB could not be observed at all.

The design of the first generation device was relatively complex, with possible spurious modes and couplings between the CPB, nanoresonator, LC and feedline that could give rise to the additional structure in the absorption spectrum. Thus, after the second device failed to function, it was decided to implement a new design in which the CPB-coupled nanoresonator was embedded within the ground plane of a CPW cavity. As well, to reduce the possible influence of charge-based fluctuators and TLS at high voltages, it was decided to engineer the CPB in the transmon regime.  These changes were implemented in Generation II and III as discussed in the following two sections.          

 
\subsection{Generation II: CPB Integrated with CPW Cavity and Flexural Nanoresonator} 
\begin{figure}
\begin{center}\includegraphics[ width=.95\columnwidth,keepaspectratio]{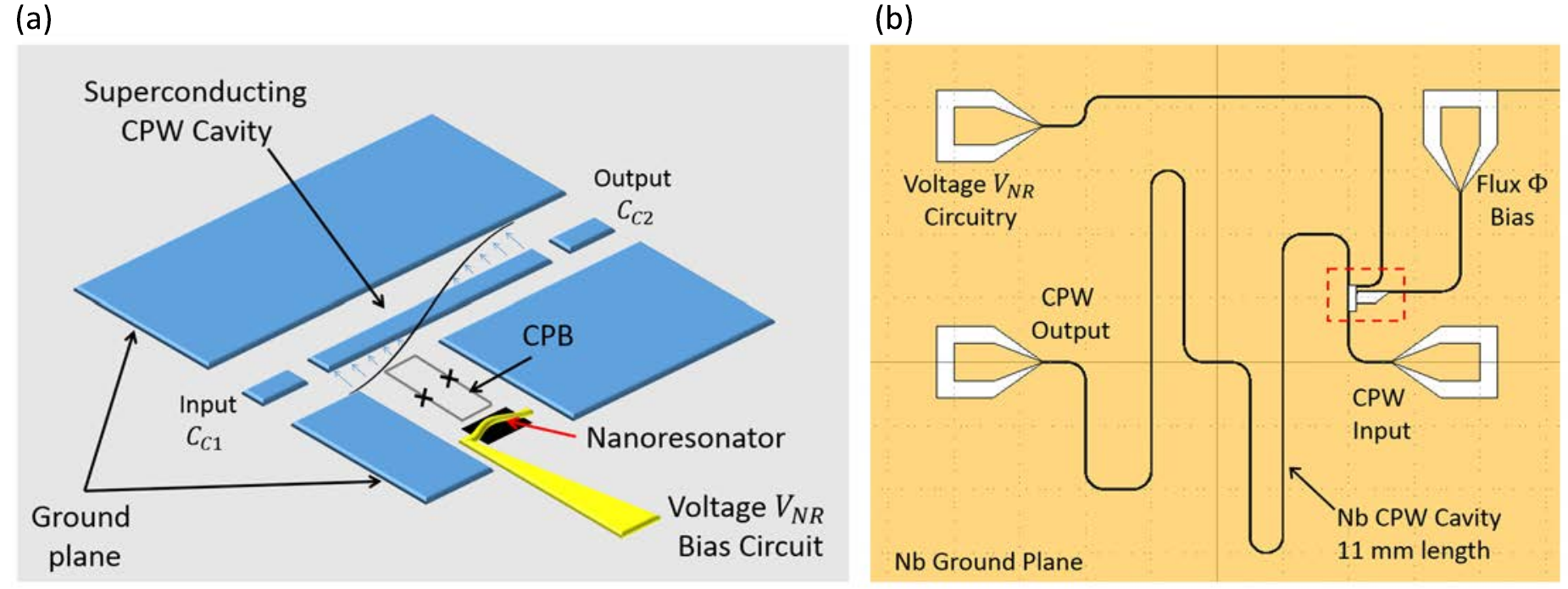}\end{center}
	\caption{General schematic and circuit design for integrating a CPB within a superconducting CPW cavity. (a) In this schematic, the CPB-coupled nanoresonator is embedded in a pocket in the ground plane of the CPW cavity, and the pocket is located in the vicinity of a voltage anti-node of the fundamental mode.  The cavity serves both for read-out and electromagnetic filtering of the CPB and nanoresonator. (b) In Generation II devices, the input and output ports, as well as all the bias lines (for $V_{\mathit{NR}}$ and $\Phi$), were fed by $50\,\Omega$ transmission lines that tapered to bond pads for wiring to external circuitry.}
	\label{fig:fig7}
\end{figure}

\begin{figure}
\begin{center}\includegraphics[ width=.95\columnwidth,keepaspectratio]{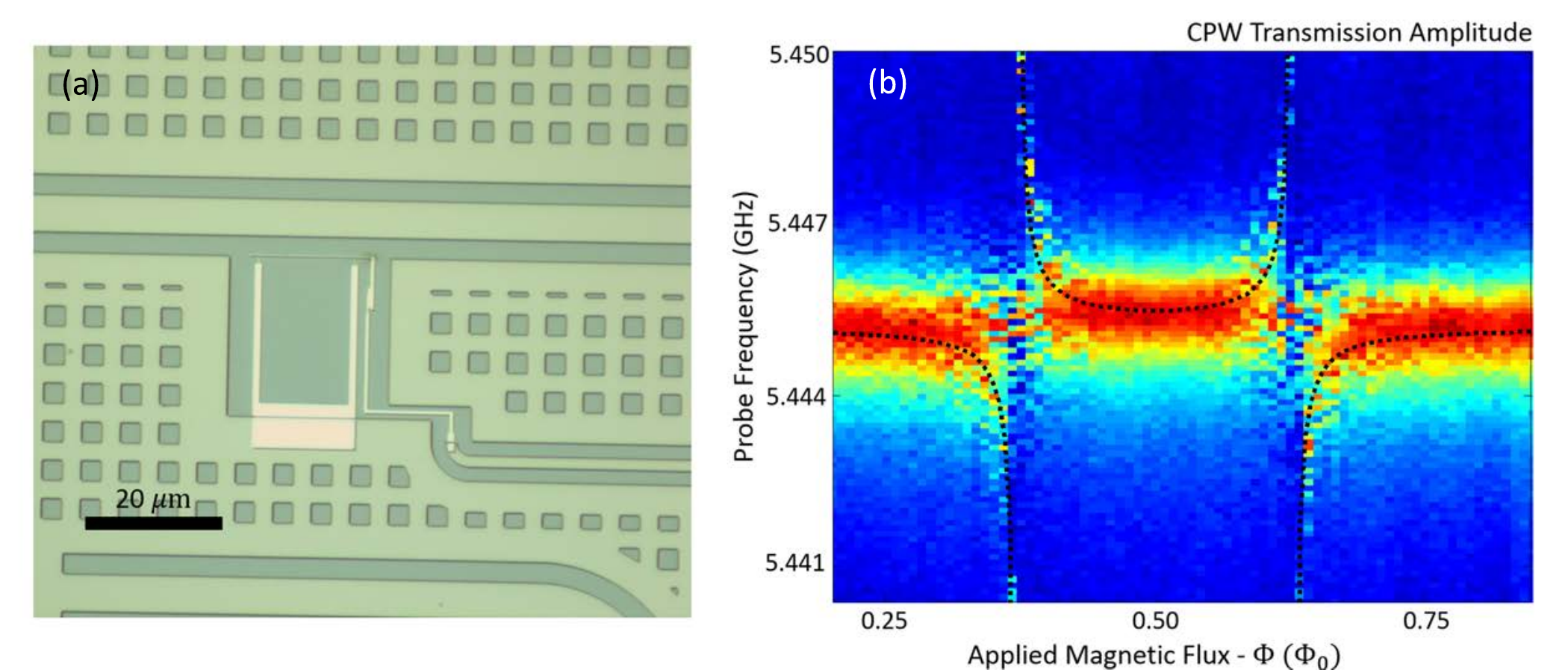}\end{center}
	\caption{Optical image and single-tone spectroscopy data for a CPB integrated within a superconducting CPW cavity. (a) Optical image of a sample from Generation II displaying the CPB, unsuspended nanoresonator electrode, flux bias line, and CPW center trace.  (b) In single-tone spectroscopy measurements of the CPW as a function of $\Phi$, periodic avoided level crossings are seen at values of flux where $\Delta E_{\mathit{CPB}}\approx\hbar\omega_{cpw}$ and are indicative of the usual hybridization of the two systems' energy bands.  Due to the low quality factor of the CPW cavity mode that resulted from parasitic losses through the nanoresonator electrode, the avoided level crossings were somewhat blurred. The dashed lines are plots of the lowest two transition energies of the coupled CPW-CPB system using numerical calculations and the following parameters:  $E_C/h=3$ GHz, $E_{J0}/h=10$ GHz, and $g/h=50$ MHz.}
	\label{fig:fig8}
\end{figure}

The second generation of devices (Figs. \ref{fig:fig7} and \ref{fig:fig8}) was developed and measured in 2013. They featured one key difference from Generation I: the CPB and nanoresonator were embedded in a superconducting CPW cavity instead of a low-frequency lumped-element LC circuit.  The CPW cavity was to play the same role as the LC in the first generation, providing read-out and isolation of the CPB-coupled nanoresonator.  However, the CPW design had the additional benefit of a much simpler mode spectrum and reduced parasitic couplings in comparison with the large LC circuit and feedline from Generation I; there was a wealth of information in the literature on the characteristics of superconducting CPWs\cite{goppl2008coplanar} and thus the transmission properties could be readily understood and modeled both analytically and numerically.

The CPW cavities consisted of a 50 $\Omega$ planar transmission line fabricated from sputtered niobium atop high resisitivity silicon substrates.  The center trace of the transmission line was 6 $\mu\rm{m}$ wide and separated by 3 $\mu\rm{m}$ on both sides from a Nb ground plane.  The cavity was formed from two gaps in the transmission line that also served as input and output coupling capcitors $C_{C1}$  and $C_{C2}$ [Fig. 7(a)] for performing transmission measurements of the cavity's frequency response.  The total length of the cavity was designed to be $\sim$11 mm [Fig.~\ref{fig:fig7}(b)], yielding fundamental mode frequencies of $\sim$ 5.4 GHz, which agreed very well with EM field-solver simulations using the commercial software Sonnet.   The coupling capacitors were designed to be symmetric with values  $C_{C1} = C_{C2}=2\,{\rm fF}$, which should have yielded a coupling quality factor $Q_{C} = 6.5 \times 10^4$.  However, the loaded quality factor $Q_L$ of the CPW fundamental mode was found to be quite low and limited to a maximum of $4 \times 10^3$ at high cavity power.  As discussed below, it was determined that $Q_L$ was limited by losses through the parasitic coupling to the NR electrode.   

As illustrated in Fig. \ref{fig:fig8}(a), the CPB and nanoresonator were fabricated in a pocket in the ground plane near one of the voltage anti-nodes of the fundamental resonance. The CPB island was arranged to be parallel with the center trace of the CPW and flush with the edge of the ground plane. For this generation, the CPB was designed to be closer to the charge qubit regime, with $E_{C}/h\approx3\,{\rm GHz}$ as determined with Q3D simulations and $E_J/h\approx10\,{\rm GHz}$.  Just like for the case of the LC circuit, the capacitive coupling $C_Q$ between the CPB and CPW center trace yields an interaction analogous to Eq. (\ref{HamiltonianINT}) with interaction strength given by Eq. (\ref{lccoupling}). For this geometry, simulations calculated $C_Q\approx 0.8\,{\rm fF}$, which would yield a coupling stength $\lambda_{CPW}/2\pi\approx 60\,{\rm MHz}$. This agreed well with single-tone spectroscopy measurements of the cavity and CPB which displayed the usual $\Phi_0$-periodic avoided level crossings  at values of $\Phi$ where the CPB and CPW were in resonance [Fig. \ref{fig:fig8}(b)].  Here the currents for tuning $\Phi$ were applied through a $50\,\Omega$ Nb trace on chip that was set back in the ground plane $\sim 25\,\mu{\rm m}$ from the CPB.

The nanoresonator electrode was fabricated 80 to 100 nm from the CPB island (not shown).  It was connected to a $50\,\Omega$ Nb trace that meandered through the ground plane and eventually tapered to a bond pad so that connections could be made to supply the coupling voltage $V_{\mathit{NR}}$. Two-tone spectroscopy measurements versus $V_{\mathit{NR}}$ suggested that the coupling between the CPB and full-length nanoresonator electrode was actually quite large, $C_{\mathit{NR}} \sim 1\,{\rm fF}$.  However, simulations were not done to determine how much of $C_{\mathit{NR}}$ was contributed from the portion of the electrode that was to be suspended to form the actual nanoresonator.  In fact, the nanoresonator was never suspended for measurements with this generation.  This was the case because preliminary single-tone spectroscopy measurements of the CPB and CPW with the resonator unetched showed that the CPW was heavily damped [Fig. 8(b)]. Further investigation with Q3D simulations showed that the parasitic capacitance between the trace to the nanoresonator electrode and the CPW center could explain the excess loading of the cavity quality factor.  Moreover, simulations using Sonnet also illustrated that a significant fraction of the cavity signal was transmitted to the nanoresonator lead.  As a result of this, measurements were stopped prematurely in order to redesign the samples to introduce $V_{\mathit{NR}}$ without degrading the cavity (or CPB) quality.   


\subsection{Generation III: CPB in the Transmon Regime Integrated with CPW Cavity, Flexural Nanoresonator, and Superconducting T-filter} 
\label{subsection:sub3}
In 2014, to overcome the excessive cavity damping observed in Generation II devices due to parasitic coupling to the nanoresonator DC bias circuitry, the LaHaye group developed a new superconducting microwave filter that can be integrated with cQED architectures to apply DC biases without degrading CPW cavity mode quality.\cite{hao2014development}  As described in in Ref. \citen{hao2014development}, the filter design utilizes on-chip, planar meander inductors and inter-digitated capacitors to form a reflective \textit{t-filter} that strongly attenuates ($\sim25\,{\rm dB}$) signals in the range from $2\,{\rm GHz}$ to $10\,{\rm GHz}$. Importantly, it was shown that the filter could be integrated into a CPW cavity [Fig. 9(a)] allowing for application of DC voltages without distorting the frequency response or reducing $Q_{L}$ of the fundamental mode, even for quality factors as high as $Q_L=2\times10^5$ and voltages as large as $V_{\mathit{NR}}=20\,{\rm V}$.\cite{hao2014development}  
 
\begin{figure}
\begin{center}\includegraphics[ width=.95\columnwidth,keepaspectratio]{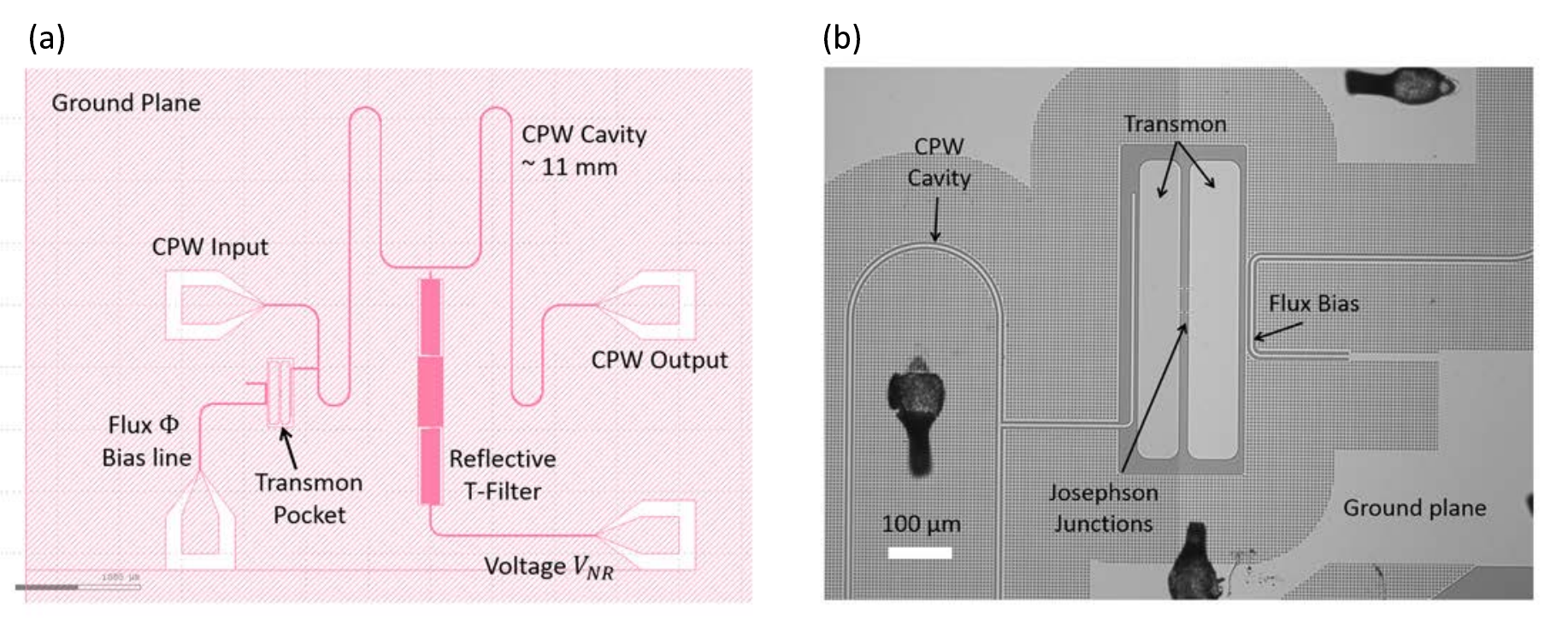}\end{center}
	\caption{Design and SEM micrograph of a transmon integrated in a voltage-biased CPW cavity for Generation III devices. (a) To eliminate the problem of excess cavity damping due to the introduction of DC bias circuitry into the cavity, a microwave t-filter was integrated with the CPW layout. $V_{\mathit{NR}}$ could then be applied to CPB and nanoresonator through this filter without degrading the CPW's fundamental mode frequency response or quality factor as discussed in Ref. \citen{hao2014development}. (b) In Generation III, CPB qubits in the transmon regime were embedded in the ground plane of the t-filtered CPW.}
	\label{fig:fig9}
\end{figure}

In subsequent and ongoing work at Syracuse, a transmon qubit was integrated with the new filtered CPW cavity design [Fig.~\ref{fig:fig9}(b)], and preliminary tests of the influence of the filter on the transmon's characteristics performed. In two-tone spectroscopic measurements, the number-state statistics of the CPW cavity\cite{schuster2007circuit} were observable, with no apparent increase in transition linewidth, for linewidths as small as $2\,{\rm MHz}$ and voltages as large as $V_{\mathit{NR}}=8V$ (not shown).  More recently, time-domain measurements of a similar transmon have been made using dispersive read-out with the t-filtered cavity.  Both Rabi oscillations and relaxation measurements were performed at $V_{\mathit{NR}}=0V$, from which estimates of $T_2^*\geq 0.5 \mu s$ and $T_1\geq12\mu s$ were obtained.    Measurements are currently underway to observe how $T_1$ and $T_2^*$  change with $V_{\mathit{NR}}$. As well, suspended nanoresonators have now been integrated with the latest samples.      

\section{Conclusions} 
The technical difficulties that were brought to light in the previous sections related to integrated nanomechanical elements, superconducting devices and circuitry like the CPW and CPB will soon be overcome, enabling a series of important experiments to probe fundamental topics such as entanglement, decoherence, and quantum measurement in new macroscopic limits.  The dispersive measurement techniques that are developed will also pave the way not only for generating Schr\"{o}dinger-cat states of mechanical structures but also for quantum non-demolition measurements of the energy of such structures, potentially allowing for new studies of energy transfer and dissipation at the mesoscale.  These systems will also play important roles in the revolution of engineered quantum systems that is now beginning, serving as elements in quantum information and communication architectures and components in quantum sensing technologies.  To continue the development further into the future, a new set of challenges arises:  integrating superconducting quantum electromechanical systems with optical technology; interfacing superconducting devices like the CPB with truly macroscopic systems (beyond the nano and micromechanical regimes); and developing these systems for an array of sensing applications. 

\acknowledgments     
 
The authors would like to thank B. Plourde for technical assistance and helpful conversations. The work was performed in part at the Cornell NanoScale Facility, a member of the National Nanotechnology Infrastructure Network, which is supported by the National Science Foundation (Grant ECCS-0335765). The authors acknowledge support for this work provided by the National Science Foundation under Grant DMR-1056423 and Grant DMR-1312421


\bibliographystyle{spiebib}   
\bibliography{masterbibfile2}   

\end{document}